\documentstyle[preprint,array,aps]{revtex}
\begin{document}
\draft
\title
{Lorentz covariant reduced-density-operator theory for
relativistic quantum information processing}
\author{Doyeol Ahn$^{1,2}${\footnote{e-mail:dahn@uoscc.uos.ac.kr}},
Hyuk-jae Lee$^1${\footnote{e-mail:lhjae@iquips.uos.ac.kr}}and Sung
Woo Hwang$^{1,3}${\footnote{e-mail:swhwang@korea.ac.kr}}}
\address{
$^1$Institute of Quantum Information Processing and Systems,
University of Seoul, Seoul, 130-743, Korea\\
$^2$Department of Electrical and Computer Engineering, University
of Seoul,
Seoul, 130-743, Korea\\
$^3$Department of Electronic Engineering, Korea University, Seoul,
136-701, Korea}

\maketitle


\baselineskip24pt



\begin{abstract}
In this paper, we derived Lorentz covariant quantum Liouville
equation for the density operator which describes the relativistic
quantum information processing from Tomonaga-Schwinger equation
and an exact formal solution for the reduced-density-operator is
obtained using the projector operator technique and the functional
calculus. When all the members of the family of the hypersurfaces
become flat hyperplanes, it is shown that our results agree with
those of non-relativistic case which is valid only in some
specified reference frame. To show that our new formulation can be
applied to practical problems, we derived the polarization of the
vacuum in quantum electrodynamics up to the second order. The
formulation presented in this work is general and could be applied
to related fields such as quantum electrodynamics and relativistic
statistical mechanics.
\end{abstract}
\vspace{.25in}


\newpage

Recently, there has been growing interest in the relativistic
formulation \cite{ahar}-\cite{ahn} of quantum operations for possible near future
applications to relativistic quantum information processing such
as teleportation \cite{benn}, entanglement-enhanced communication \cite{benne}, and quantum
clock synchronization \cite{jozs}, \cite{chua}.

In the non-relativistic case, the key element for studying quantum
information processing is the density operator of a quantum register
which is derived from the solution of a quantum Liouville equation
(QLE)\cite{ahnd}, \cite{ahndo} for the total system including an environment. The QLE is an
integro-differential equation and it is in general nontrivial to
obtain the solution of the form
\begin{equation}
\rho\stackrel{\cal E}{\rightarrow}\rho'=\hat{\cal E}[\rho],
\label{dens}
\end{equation}
where $\rho$ is the reduced density operator of the quantum register and $\hat{\cal E}$ is
the superoperator describing the evolution of $\rho$ by the quantum information processing. In the previous
works, we have employed a time-convolutionless reduced-density-operator formalism to
model quantum devices \cite{ahndoy} and noisy quantum channels \cite{ahni},\cite{ahnii}.

The first step toward the relativistic quantum information theory
would be the formulation of Lorentz covariant QLE and the
derivation of the reduced-density-operator which is a solution of
the covariant QLE. The goal of this paper is to derive Lorentz
covariant quantum Liouville equation which describes the
relativistic quantum information processing and obtain a formal
solution for the reduced-density-operator pertaining to the system
(or electrons) part alone.

It is well known that neither the non-relativistic Schr\"{o}dinger equation nor the
QLE is Lorentz covariant. As a result, it is expected that the usual non-relativistic definition of the
reduced-density-operator and its functionals such as quantum entropy have no invariant meaning in special
relativity. Another conceptual barrier for the relativistic treatment of quantum information processing
is the difference of the role played by the wave fields and the state vectors in the quantum field theory.
In non-relativistic quantum mechanics both the wave function and the state vector in Hilbert space give
the probability amplitude which can be used to define conserved positive probability densities or density
matrix. On the other hands, in relativistic quantum field theory, covariant wave fields are not probability
amplitude at all, but operators which create or destroy particles in spanned by states defined as containing
definite numbers of particles or antiparticles in each normal mode \cite{wein}. The role of the fields is to make
the interaction or S-matrix satisfy the Lorentz invariance and the cluster decomposition principle. The
information of the particle states is contained in the state vectors of the Hilbert space spanned by states
containing $0, 1, 2, \cdots$ particles as in the case of non-relativistic quantum mechanics. So it seems
like that one needs to obtain the covariant equation of motion for the state vector and derive the
covariant QLE out of it.

Some time ago, Tomonaga \cite{tomo} and Schwinger \cite{schw}
derived a covariant equation of motion for the quantum state
vector in terms of the functional derivative, known as
Tomonaga-Schwinger (T-S) equation,
\begin{equation}
i\frac{\delta\Psi[\sigma]}{\delta\sigma(x)}={\cal H}_{int}(x)\Psi[\sigma],
\label{func}
\end{equation}
in the interaction picture. Here $x$ is a space-time four-vector, $\sigma$ is the spacelike hypersurface, $\Psi[\sigma]$ is the state vector
which is a functional of $\sigma$, ${\cal H}_{int}(x)={\cal H}_{int}[\varphi_{\alpha}(x)]$ is the interaction
Hamiltonian density which is a functional of quantum field $\varphi_{\alpha}[x]$, and $\frac{\delta}{\delta\sigma(x)}$ is
the Lorentz invariant functional derivative \cite{nish}. The functional derivative of $\Psi[\sigma]$ is defined as
\begin{equation}
\frac{\delta\Psi[\sigma]}{\delta\sigma(x)}=\lim_{\delta\omega\to
0}\frac{\Psi[\sigma']-\Psi[\sigma]}{\delta\omega},
\label{funci}
\end{equation}
where $\delta\omega$ is an infinitesimal four-dimensional volume between two hypersurfaces $\sigma$ and $\sigma'$.
The formal solution of equation (\ref{func}) is given by
\begin{equation}
\Psi[\sigma]={\cal U}[\sigma, \sigma_0]\Psi[\sigma_0],
\label{wave}
\end{equation}
where the generalized transformational functional satisfies the T-S equation
\begin{equation}
i\frac{\delta{\cal U}[\sigma, \sigma_0]}{\delta\sigma(x)}={\cal H}_{int}(x){\cal U}[\sigma,\sigma_0]
\label{tseq}
\end{equation}
with the boundary condition ${\cal U}[\sigma_0, \sigma_0]=1$. The generalized transformation functional ${\cal U}[\sigma, \sigma_0]$
is a unitary operator. We also have \cite{schw}
\begin{equation}
\frac{\delta{\cal U}^{-1}[\sigma, \sigma_0]}{\delta\sigma(x)}=-{\cal U}^{-1}[\sigma, \sigma_0]
\frac{\delta{\cal U}[\sigma, \sigma_0]}{\delta\sigma(x)}{\cal U}^{-1}[\sigma, \sigma_0],
\label{utra}
\end{equation}
from the unitary condition. Throughout the paper, we assume $\hbar=c=1$. The expectation value of some field variable $F(x)$
becomes
\begin{eqnarray}
<F(x)>&=&(\Psi[\sigma], F(x)\Psi[\sigma])\nonumber\\
&=&trace(F(x)\Psi[\sigma]\Psi^{\dag}[\sigma])\nonumber\\
&=&trace(F(x)\rho_T[\sigma]).
\label{expe}
\end{eqnarray}
>From equation (\ref{expe}), we notice that the total density operator $\rho_T[\sigma]$ can be written as \cite{maff}\cite{holl}
\begin{eqnarray}
\rho_T[\sigma]&=&\Psi[\sigma]\Psi^{\dag}[\sigma]\nonumber\\
&=&{\cal U}[\sigma, \sigma_0]\Psi[\sigma_0]\Psi^{\dag}[\sigma_0]{\cal U}^{-1}[\sigma, \sigma_0].
\label{tden}
\end{eqnarray}
Then,
\begin{eqnarray}
\frac{\delta\rho_T[\sigma]}{\delta\sigma(x)}&=&\frac{\delta}{\delta\sigma}\{{\cal U}[\sigma, \sigma_0]
\Psi[\sigma_0]\Psi^{\dag}[\sigma_0]{\cal U}^{-1}[\sigma, \sigma_0]\}\nonumber\\
&=&[\frac{\delta{\cal U}[\sigma, \sigma_0]}{\delta\sigma(x)}{\cal U}^{-1}[\sigma, \sigma_0], \rho_T[\sigma]]\nonumber\\
&=&-i [{\cal H}_{int}(x), \rho_T[\sigma]]\nonumber\\
&=&-i\hat{\cal L}(x)\rho_T[\sigma],
\label{liou}
\end{eqnarray}
where $\hat{\cal L}(x)$ is the Liouville superoperator. Since equation (\ref{liou}) describes the Lorentz covariant equation
of motion for the total density operator, we denote it as the covariant quantum Liouville equation (CQLE). Note that the Liouville
superoperator is not an operator in the Hilbert space of state vectors but a linear operator in the Hilbert-Schmidt space of density
matrices \cite{ahnii}. Here $\rho_T[\sigma]$ contains the information for the total system, for example, an interacting spin-$\frac{1}{2}$ massive
particles and photons in the case of quantum electrodynamics (QED).

In order to extract the information of the system or the electrons alone, it is convenient to use
the projection operators \cite{ahnd}, \cite{zwan}, \cite{saek} that decompose
the total system by eliminating the degrees of freedom for the environment, say, the photon field in the case of QED.
The information of the system is then contained in the reduced-density-operator $\rho[\sigma]$ which is defined as
\begin{eqnarray}
\rho[\sigma]&=&tr_B\rho_T[\sigma]\nonumber\\
&=&tr_B{\cal P}\rho_T[\sigma],
\label{rden}
\end{eqnarray}
where the projection operator ${\cal P}$ and ${\cal Q}$ are defined as
${\cal P}X=\rho_B tr_B(X),  {\cal Q}=1-{\cal P}$,
for any covariant dynamical variable $X$, $\rho_B$ is the density matrix for the quantum environment at $\sigma_0$ and $tr_B$ indicates a partial
trace over the quantum environment. The projection operators satisfy the operator identities ${\cal P}^2={\cal P}, {\cal Q}^2={\cal Q}$,
${\cal P}{\cal Q}={\cal Q}{\cal P}=0$ and $[\frac{\delta}{\delta\sigma(x)}, {\cal P}]=[\frac{\delta}{\delta\sigma(x)}, {\cal Q}]=0$.
Furthermore, we would like to note that  $(\frac{\delta}{\delta\sigma(x)})^{-1}=\int d^4x$ \cite{nish}, and the system and the
environment are decoupled at $\sigma_0$. We also note that the projection operators ${\cal P}$ and ${\cal Q}$
are functionals of the initial hypersurface $\sigma_0$($\neq \sigma$ for all $x$) and unless otherwise specified, we will
omit the functional argument. However, one needs to keep track of the functional argument especially in the four-dimensional
integration.

The CQLE (\ref{liou}) can be decomposed into two coupled equation for ${\cal P}\rho_T[\sigma]$ and ${\cal Q}\rho_T[\sigma]$:
\begin{mathletters}
\begin{equation}
\frac{\delta}{\delta\sigma(x)}{\cal P}\rho_T[\sigma]=-i{\cal P}\hat{\cal L}(x){\cal P}\rho_T[\sigma]-i{\cal P}\hat{\cal L}(x){\cal Q}\rho_T[\sigma],
\label{pP}
\end{equation}
\begin{equation}
\frac{\delta}{\delta\sigma(x)}{\cal Q}\rho_T[\sigma]=-i{\cal Q}\hat{\cal L}(x){\cal Q}\rho_T[\sigma]-i{\cal Q}\hat{\cal L}(x){\cal P}\rho_T[\sigma],
\label{pQ}
\end{equation}
\end{mathletters}

In order to obtain the formal solution, we solve first eq. (\ref{pQ}) using the integrating factor.
Let $h[\sigma]$ be an integrating factor such that
\begin{eqnarray}
h[\sigma]\{\frac{\delta}{\delta\sigma(x)}{\cal Q}\rho_T[\sigma]+i {\cal Q}\hat{\cal L}(x){\cal Q}\rho_T[\sigma]\}&=&
-i h[\sigma]{\cal Q}\hat{\cal L}(x){\cal Q}\rho[\sigma]\nonumber\\
&=&\frac{\delta}{\delta \sigma(x)}\{h[\sigma]{\cal Q}\rho[\sigma]\}.
\label{infac}
\end{eqnarray}
Then, $\frac{\delta h[\sigma]}{\delta\sigma(x)}=i h[\sigma]{\cal Q}\hat{\cal L}(x){\cal Q}$ and we obtain
\begin{equation}
h[\sigma]=T^c \exp\{ i\int^{\sigma}_{\sigma_0} d^4 x'{\cal Q}\hat{\cal L}(x'){\cal Q}\}.
\label{factO}
\end{equation}
>From eq. (\ref{infac}),
\begin{eqnarray}
{\cal Q}\rho_T [\sigma] &=& h^{-1}[\sigma]h[\sigma_0]{\cal Q}\rho[\sigma_0]-
i\int^{\sigma}_{\sigma_0} d^4 x' h^{-1}[\sigma(x)]h[\sigma (x')]{\cal Q}\hat{\cal L}(x'){\cal P}\rho_T [\sigma(x')]\nonumber\\
&=&-i\int^{\sigma}_{\sigma_0} d^4 x' H[\sigma(x),\sigma (x')]{\cal Q}\hat{\cal L}(x'){\cal P}\rho_T [\sigma(x')],
\label{Qint}
\end{eqnarray}
where we assume that $\rho_T[\sigma]$ is decoupled when $\sigma=\sigma_0$ and
\begin{equation}
\hat{H}[\sigma(x), \sigma(x')]=T\exp\{-i\int^{\sigma(x)}_{\sigma(x')} d^4 x'' {\cal Q}\hat{\cal L}(x''){\cal Q}\}.
\end{equation}
Here T and  $T^c$ are time-ordering and anti-time ordering operators, respectively, and $H[\sigma(x),\sigma (x')]$
is the projected propagator. In order to derive the convolutionless equation of motion, we define the retarded propagator
$G_R [\sigma(x), \sigma(x')]$ such that
\begin{equation}
\hat{G}_R[\sigma(x),\sigma(x')]=T^c \exp \{i\int^{\sigma(x)}_{\sigma(x')} d^4 x'' \hat{\cal L}(x'')\}
\end{equation}
which satisfies
\begin{equation}
\rho_T [\sigma_0]=\hat{G}_R [\sigma, \sigma_0]\rho[\sigma].
\end{equation}
Then,
\begin{eqnarray}
{\cal Q}\rho[\sigma]&=&-i\int^{\sigma}_{\sigma_0} d^4 x'
H[\sigma(x),\sigma (x')]{\cal Q}\hat{\cal L}(x'){\cal P}\hat{G}_R[\sigma(x),\sigma(x')]\rho_T [\sigma(x)]\nonumber\\
&=&-i\int^{\sigma}_{\sigma_0} d^4 x'
H[\sigma(x),\sigma (x')]{\cal Q}\hat{\cal L}(x'){\cal P}\hat{G}_R[\sigma(x),\sigma(x')]{\cal P}\rho_T [\sigma(x)]\nonumber\\
&&-i\int^{\sigma}_{\sigma_0} d^4 x'
H[\sigma(x),\sigma (x')]{\cal Q}\hat{\cal L}(x'){\cal P}\hat{G}_R[\sigma(x),\sigma(x')]{\cal Q}\rho_T [\sigma(x)]
\label{qrho}
\end{eqnarray}
and
\begin{equation}
{\cal Q}\rho_T[\sigma]=\{\theta[\sigma]-1\}{\cal P}\rho_T[\sigma],
\label{pQso}
\end{equation}
where
\begin{equation}
\theta^{-1}[\sigma]=1+i\int^{\sigma}_{\sigma_0}d^4x'H[\sigma(x),\sigma(x')]{\cal Q}\hat{\cal L}(x'){\cal P}G_R[\sigma(x), \sigma(x')].
\label{thet}
\end{equation}
Once the solution for ${\cal Q}\rho_T[\sigma]$ is obtained, it is substituted for the equation for ${\cal P}\rho_T[\sigma]$.
Then, after, some mathematical manipulations, we obtain using the integrating factor technique again,
\begin{equation}
{\cal P}\rho_T[\sigma]=W^{-1}[\sigma, \sigma_0]\hat{\cal U}_s[\sigma, \sigma_0]{\cal P}\rho_T [\sigma_0],
\label{prot}
\end{equation}
or
\begin{equation}
\rho[\sigma]=tr_B\{W^{-1}[\sigma, \sigma_0]\hat{\cal U}_s[\sigma, \sigma_0]\rho_B\}\rho[\sigma_0],
\label{proti}
\end{equation}
where
\begin{equation}
W[\sigma, \sigma_0]=1+i\int^{\sigma}_{\sigma_0}d^4x'\hat{\cal U}_s[\sigma(x), \sigma(x')]{\cal P}\hat{\cal L}(x')
\{\theta[\sigma(x')]-1\}{\cal P}G_R[\sigma(x), \sigma(x')]\theta[\sigma(x)],
\label{eqW}
\end{equation}
and
\begin{equation}
\hat{\cal U}_s[\sigma, \sigma_0]=T\exp\{-i\int^{\sigma}_{\sigma_0}d^4x' {\cal P}\hat{\cal L}(x'){\cal P}\}.
\label{eqU}
\end{equation}
Here $\hat{\cal U}_s[\sigma, \sigma_0]$ is the generalized transformation functional or the propagator for the reduced system.

It is remarkable to note when hypersurfaces $\sigma_0$ and all the
members of the family $\{\sigma\}$ are hyperplane flat surfaces
parametrized by $t=constant$ \cite{nish}, then the transformation
functional such as ${\cal U}_s[\sigma(x), \sigma(x')]$ can be
written as ${\cal U}_s(t,t')$. Then, if we set $t_0=0$,
\begin{eqnarray}
W(t,0)&=&W(t)\nonumber\\
&=&1+\int^{t}_{0} d^4 x' \hat{\cal U}_s (t,t'){\cal P}{\cal L}(x',t)\{\theta(t')-1\}{\cal P} G_R(t,t')\theta(t)\nonumber\\
&=&1+\int^{t}_{0} ds \hat{\cal U}_s (t,s)tr_B[{\cal L}(s)\{\theta(s)-1\}\rho_B]tr_B[G_R(t,s)\theta(t)\rho_B],
\label{wt}
\end{eqnarray}
with $\hat{\cal L}(s)=\int d^3 x'(x',s)$.
As a result, the covariant forms of equations (\ref{prot}) to
(\ref{eqU}) become reduced to those of the non-relativistic case
which is valid only in some specified reference frame given by
equations (18) to (25) of reference \cite{ahni}.

By comparing, equations (\ref{dens}) and (\ref{proti}), the covariant superoperator for the relativistic quantum operation
$\hat{\cal E}[\sigma, \sigma_0]$ can be written as
\begin{equation}
\hat{\cal E}[\sigma, \sigma_0]=tr_B\{W^{-1}[\sigma, \sigma_0]\hat{\cal U}_s[\sigma, \sigma_0]\rho_B\}.
\label{supe}
\end{equation}
So far all our results are exact and the equations (\ref{prot}) to (\ref{supe}) would be the key steps in the analysis
of relativistic quantum information processing. Apart from describing quantum information
processing, QLE and reduced-density-operator have been essential in solving various quantum optics and
non-Markovian optical problems in the non-relativistic domain \cite{ahnd}-\cite{ahndoy}. So it might be interesting to extend
this approach to revisit relativistic quantum electrodynamics problems, which were solved relying on
renormalization procedures in field theory, using the covariant form of quantum Liouville equation. On the other hand,
relativistic thermodynamics or statistical mechanics look like an area where the knowledge of the density operator or
the reduced-density-operator might come in handy provided the ambiguity of the temperature concept in special relativity
is resolved. We believe our formalism is general and could be applied to
related fields such as QED and relativistic statistical mechanics.
As a matter of fact, these related fields would also play an important role in relativistic quantum information processing
because these processes would cause the decoherence as in the non-relativistic case.

To show how to apply the formalism we developed to practical problems, we give a derivation of the polarization of the vacuum by an external field starting from the equation
(\ref{proti}). The Hamiltonian for the coupling between an electron and electromagnetic fields is given by
\begin{equation}
\hat{\cal H}(x)=-\hat{j}_{\mu}(x)\hat{A}_{\mu}(x),
\end{equation}
where $\hat{j}_{\mu}(x)$ and $\hat{A}_{\mu}(x)$ are current and electromagnetic 4-vector potential operators, respectively.
Then from eq. (\ref{proti}), the reduced density operator up to the first order in $\hat{\cal L}$ becomes
\begin{eqnarray}
\rho^{(1)} [\sigma]&=&
tr_B\{ 1- i\int^{\sigma}_{\sigma_0} d^4 x'{\cal P}{\cal L}(x'){\cal P}\} \rho[\sigma_0]\nonumber\\
&=&(1- i\int^{\sigma}_{\sigma_0} d^4 x' tr_B({\cal L}(x')\rho_B))\rho[\sigma_0].
\end{eqnarray}
If we set the initial hypersurface be the flat surface $\sigma_0 =-\infty$, $\rho[-\infty]=\rho_0$ and
$A_{\mu}(x)=tr_B(\hat{A}_{\mu}(x)\rho_B)$ which is a classical external field, we get
\begin{equation}
\rho^{(1)}[\sigma]=\rho_0 + i\int^{\sigma}_{-\infty} d^4 x' [\hat{j}_{\mu}(x')A_{\mu}(x'), \rho_0].
\label{fsol}
\end{equation}
The polarization of the vacuum is the expectation value of $\hat{j}_{\mu}(x)$, computed for the state of the system as
modefied by the external electromagnetic field \cite{schwi}, \cite{schwii} and is given by
\begin{eqnarray}
\langle\hat{j}_{\mu}(x)\rangle &=& tr (\hat{j}_{\mu}(x)\rho^{(1)}[\sigma])\nonumber\\
&=&tr (\hat{j}_{\mu}(x)\rho_0)+i\int^{\sigma}_{-\infty} d^4 x' tr\{[\hat{j}_{\nu}(x')A_{\nu}(x'), \rho_0]\hat{j}_{\mu}(x)\}\nonumber\\
&=&i\int^{\sigma}_{-\infty} d^4 x' tr\{[\hat{j}_{\mu}(x), \hat{j}_{\nu}(x')]\rho_0\}A_{\nu}(x')\nonumber\\
&=&i\int^{\sigma}_{-\infty} d^4 x' \langle[\hat{j}_{\mu}(x), \hat{j}_{\nu}(x')]\rangle_0 A_{\nu}(x')\nonumber\\
&=&-\frac{\alpha}{15}\frac{1}{k_0^2}\Box^2 J_{\mu}(x)+\cdots
\label{expj}
\end{eqnarray}
where $tr (\cdots)$ is the trace over the electron states and $\langle\cdots\rangle_0$ is the expectation value for the electron fields.
Here $J_{\mu}(x)$ is the external current generating the electromagnetic field, $k_0=m_0 c/\hbar$, $\alpha=e^2/4\pi \hbar c$,
$\Box^2=\partial_{\mu}\partial^{\mu}$, and $m_0$ is the electron mass\cite{schwi}\cite{schwii}.
Eq. (\ref{expj}) describes the vacuum polarization due to the external electromagnetic fields in quantum electrodynamics.
We proceed to derive the second order correction to the vacuum polarization $\langle j_{\mu}(x)\rangle^{(2)}$. The second order
correction term to the reduced density operator $\triangle\rho^{(2)}[\sigma]$ becomes
\begin{eqnarray}
\triangle\rho^{(2)}[\sigma]&=&-i\int^{\sigma}_{-\infty} d^4 x'
tr_B\{(W^{-1}[\sigma,-\infty]\hat{\cal U}_s[\sigma, -\infty])^{(2)}\rho_B\}\rho_0\nonumber\\
&=&-2\int^{\sigma(x)}_{-\infty} d^4 x' \int^{\sigma(x')}_{-\infty} d^4 x''
tr_B(\hat{\cal L}(x')\hat{\cal L}(x'')\rho_B)\rho_0\nonumber\\
&=&-2\int^{\sigma(x)}_{-\infty} d^4 x' \int^{\sigma(x')}_{-\infty} d^4 x''
\{-\langle\hat{A}_{\mu}(x')\hat{A}_{\nu}(x'')\rangle_0\hat{j}_{\mu}(x')\hat{j}_{\nu}(x'')\rho_0\nonumber\\
&&+\langle\hat{A}_{\mu}(x')\hat{A}_{\nu}(x'')\rangle_0\hat{j}_{\nu}(x'')\rho_0\hat{j}_{\mu}(x')
+\langle\hat{A}_{\nu}(x'')\hat{A}_{\mu}(x')\rangle_0\hat{j}_{\mu}(x')\rho_0\hat{j}_{\nu}(x'') \nonumber\\
&&-\langle\hat{A}_{\nu}(x'')\hat{A}_{\mu}(x')\rangle_0\rho_0\hat{j}_{\nu}(x'')\hat{j}_{\mu}(x')\}.
\label{sden}
\end{eqnarray}
Then,
\begin{eqnarray}
\langle\hat{j}_{\mu}(x)\rangle^{(2)}&=& tr(\hat{j}_{\mu}(x)\triangle\rho^{(2)}[\sigma])\nonumber\\
&=&-2\int^{\sigma(x)}_{-\infty} d^4 x' \int^{\sigma(x')}_{-\infty} d^4 x''
\{-\langle\hat{A}_{\mu}(x')\hat{A}_{\nu}(x'')\rangle_0\langle[\hat{j}_{\lambda}(x),\hat{j}_{\mu}(x')] \hat{j}_{\nu}(x'')\rangle_0\nonumber\\
&&+\langle\hat{A}_{\nu}(x'')\hat{A}_{\mu}(x')\rangle_0\langle\hat{j}_{\nu}(x'')[\hat{j}_{\lambda}(x),\hat{j}_{\mu}(x')] \rangle_0\}\nonumber\\
&=&-2\int^{\sigma(x)}_{-\infty} d^4 x' \int^{\sigma(x')}_{-\infty} d^4 x''
\langle[\hat{A}_{\mu}(x'),\hat{A}_{\nu}(x'')\rangle_0\langle[\hat{j}_{\lambda}(x),\hat{j}_{\mu}(x')] \hat{j}_{\nu}(x'')\rangle_0\nonumber\\
&=&-2i\int^{\sigma(x)}_{-\infty} d^4 x' \int^{\sigma(x')}_{-\infty} d^4 x''
\delta_{\mu\nu}D(x'-x'')\langle[\hat{j}_{\lambda}(x),\hat{j}_{\mu}(x')] \hat{j}_{\nu}(x'')\rangle_0\nonumber\\
&=&-2i \int^{\sigma(x)}_{-\infty} d^4 x' \int^{\sigma(x')}_{-\infty} d^4 x''
\langle[\hat{j}_{\mu}(x),\hat{j}_{\nu}(x')]\hat{j}_{\nu}(x'')\rangle_0 {\cal D}(x'-x''),
\label{scur}
\end{eqnarray}
where ${\cal D}(x)$ is the invariant function defined by eq.(2.17) of reference \cite{schw}. Above result can be further simplified by
using that \cite{schwi}
\begin{eqnarray}
&&\int^{\sigma(x')}_{-\infty} d^4 x'' {\cal D}(x'-x'') \hat{j}_{\mu}(x'')\nonumber\\
&&=\int^{\infty}_{-\infty} d^4 x'' \epsilon(x',x''){\cal D}(x'-x'') \hat{j}_{\mu}(x'')\nonumber\\
&&=\int^{\infty}_{-\infty} d^4 x''  \bar{\cal D}(x'-x'') \hat{j}_{\mu}(x'')\nonumber\\
&&=-2 \delta\hat{A}_{\mu}(x')
\end{eqnarray}
Here $\delta\hat{A}_{\mu}(x)$ is the 4-vector potential induced by the polarization of the vacuum or the reaction of the virtual
electron-positron coupling. Then the vacuum polarization up to the second order interaction becomes
\begin{equation}
\langle\hat{j}_{\mu}(x)\rangle=i\int^{\sigma(x)}_{-\infty} d^4 x'\langle[\hat{j}_{\mu}(x),\hat{j}_{\nu}(x')]\rangle_0\{A_{\nu}(x')
+4\delta A_{\nu}(x')\}
\end{equation}
The knowledge of vacuum polarization would be important in understanding the decoherence process in the relativistic domain. At this stage,
we would like to leave the detailed calculations of the second and higher order corrections for the future work.

In summary, we have derived Lorentz covariant quantum Liouville equations for the density operator in functional of
hypersurface from T-S equation and obtained formal solution for the reduced-density-operator which is also in covariant form using
the projection operator technique and the functional calculus.
When all the members of the family of the hypersurfaces become flat hyperplanes, our results agree with those of the
non-relativistic case. We have shown that our new formalism can be applied to the practical cases such as the vacuum polarization.
Our formulation is exact and general so it could be applied not only to the
relativistic quantum information processing but also to the related fields such as QED, field theory and
relativistic statistical mechanics.

\vspace{1.0cm}

\centerline{\bf Acknowledgements}

This work was supported by the Korean Ministry of Science and
Technology through the Creative Research Initiatives Program under
Contact No. M1-0116-00-0008.


\begin{thebibliography}{99}
\bibitem{ahar} Y. Aharonov and D. Z. Albert, Phys. Rev. {\bf D24},
359(1981).
\bibitem{czac} M. Czachor, Phys. Rev. {\bf A55}, 72(1997).
\bibitem{pere} A. Peres, Phys. Rev. {\bf A61}, 022117(2000).
\bibitem{peres} A. Peres, P. F. Scudo and D. R. Terno, Phys. Rev. Lett. {\bf 88}, 230402(2002).
\bibitem{ging} R. M. Gingrich and C. Adami, LANL e-print, quant-ph/0205179.
\bibitem{alsi} P. M. Alsing and G. J. Milburn, Quantum Information and Computation, {\bf 2}, 487(2002)
\bibitem{ahn} D. Ahn, H.-j. Lee and S. W. Hwang, LANL e-print, quant-ph/0207018;
D. Ahn, H.-j. Lee, Y. H. Moon and S. W. Hwang, LANL e-print, quant-ph/0209164, Phys. Rev. {\bf A} in press.
\bibitem{benn} C. H. Bennett, G. Brassard, C. Cr\'{e}peau, R.
Jozsa, A. Peres and W. K. Wooters, Phys. Rev. Lett. {\bf 70},
1895(1993).
\bibitem{benne} C. H. Bennett and S. J. Wiesner, Phys. Rev. Lett. {\bf 69}, 2881(1992).
\bibitem{jozs}R. Jozsa, D. S. Abrams, J. D. Dowling and C. P.
Williams, Phys. Rev. Lett. {\bf 85}, 2010(2000).
\bibitem{chua} I. L. Chuang, Phys. Rev. Lett. {\bf 85},
2006(2000).
\bibitem{ahnd} D. Ahn, Phys. Rev. {\bf B50}, 8310(1994).
\bibitem{ahndo} D. Ahn, Phys. Rev. {\bf B51}, 2159(1995).
\bibitem{ahndoy} D. Ahn, Prog. Quantum Electron. {\bf 21}, 249(1997) and references therein.
\bibitem{ahni} D. Ahn, J. H. Oh, K. Kimm and S. W. Hwang, Phys.
Rev. {\bf A61}, 052310(2000).
\bibitem{ahnii} D. Ahn, J. Lee, M. S. Kim and S. W. Hwang, Phys. Rev. {\bf A66}, 012302(2002).
\bibitem{wein} S. Weinberg, {\it The Quantum Theory of Fields I},
(Cambridge University Press, N.Y. 1995).
\bibitem{tomo} S. Tomonaga, Prog. Theor. Phys. {\bf 1}, 27(1946).
\bibitem{schw} J. Schwinger, Phys. Rev. {\bf 74}, 1439(1948).
\bibitem{nish} K. Nishijima, {\it Fields and Particles}, (W. A. Benjamin Inc., M.A. 1969).
\bibitem{maff} J. W. Maffat, LANL e-print, quant-ph/0204151.
\bibitem{holl} A. H\"{o}ll, V. G. Morozov and G. R\"{o}pke, LANL e-print, quant-ph/0208083.
\bibitem{zwan} R. Zwangzig, J. Chem. Phys. {\bf 33}, 1338(1960).
\bibitem{saek} M. Saeki, Prog. Theor. Phys. {\bf 67}, 1313(1982).
\bibitem{schwi} J. Schwinger, Phys. Rev. {\bf 75}, 651(1949).
\bibitem{schwii} J. Schwinger, Phys. Rev. {\bf 75}, 790(1949).


\end{thebibliography}
\end{document}